\documentclass[screen, sigconf]{acmart}

\acmYear{2025}\copyrightyear{2025}
\acmConference[ACM FAccT '25]{ACM Conference on Fairness, Accountability, and Transparency}{June 23--26, 2025}{Athens, Greece}
\acmBooktitle{ACM Conference on Fairness, Accountability, and Transparency (ACM FAccT '25), June 23--26, 2025, Athens, Greece}
\acmDOI{10.1145/3715275.3732005}
\acmISBN{979-8-4007-1482-5/25/06}

\usepackage{array}
\usepackage{multirow}
\usepackage{graphicx}
\usepackage{geometry} 

\copyrightyear{2025}
\acmYear{2025}
\setcopyright{cc}
\setcctype{by}
\acmConference[FAccT '25]{The 2025 ACM Conference on Fairness,
Accountability, and Transparency}{June 23--26, 2025}{Athens, Greece}
\acmBooktitle{The 2025 ACM Conference on Fairness, Accountability, and Transparency (FAccT '25), June 23--26, 2025, Athens,
Greece}
\acmDOI{10.1145/3715275.3732005}
\acmISBN{979-8-4007-1482-5/2025/06} 

\begin{document}

\title{Examining the Expanding Role of Synthetic Data Throughout the AI Development Pipeline}


\author{Shivani Kapania}
\authornote{This work was done when the author was an intern at Microsoft Research.}
\affiliation{%
  \institution{Carnegie Mellon University}
  \city{Pittsburgh, PA}
  \country{USA}}
\email{skapania@andrew.cmu.edu}

\author{Stephanie Ballard}
\affiliation{%
  \institution{Microsoft}
  \city{Redmond, WA}
  \country{USA}
}
\email{stephballard@microsoft.com}

\author{Alex Kessler}
\affiliation{%
  \institution{Microsoft}
  \city{Redmond, WA}
  \country{USA}
}
\email{alexkessler@microsoft.com}

\author{Jennifer Wortman Vaughan}
\affiliation{%
  \institution{Microsoft}
    \city{New York, NY}
  \country{USA}
}
\email{jenn@microsoft.com}

\renewcommand{\shortauthors}{Kapania et al.}

\raggedbottom
\newcommand{\etal}{et al. }
\newcommand{\eg}{e.g., }
\newcommand{\ie}{i.e., }

\begin{abstract}

  Alongside the growth of generative AI, we are witnessing a surge in the use of synthetic data across all stages of the AI development pipeline. It is now common practice for researchers and practitioners to use one large generative model (which we refer to as an \emph{auxiliary model}) to generate synthetic data that is used to train or evaluate another, reconfiguring AI workflows and reshaping the very nature of data.
  While scholars have raised concerns over the risks of synthetic data, policy guidance and best practices for its responsible use have not kept up with these rapidly evolving industry trends, in part because we lack a clear picture of current practices and challenges. Our work aims to address this gap. Through 29 interviews with AI practitioners and responsible AI experts, we examine the expanding role of synthetic data in AI development.
  Our findings reveal how auxiliary models are now widely used across the AI development pipeline. 
  Practitioners describe synthetic data as crucial for addressing data scarcity and providing a competitive edge, noting that evaluation of generative AI systems at scale would be infeasible without auxiliary models.
 However, they face challenges controlling the outputs of auxiliary models, generating data that accurately depict underrepresented groups, and scaling data validation practices that are based primarily on manual inspection. We detail general limitations of and ethical considerations for synthetic data and conclude with a proposal of concrete steps towards the development of best practices for its responsible use.

\end{abstract}

\maketitle


\section{Introduction}

Data is a critical building block of AI systems \cite{halevy2009unreasonable}. As the tech industry shifts attention and resources towards building large-scale, data-hungry generative models, traditional data sources are being exhausted, giving rise to an intense search for more data~\cite{nytimes1, nytimes2}.  However, developing high-quality datasets remains complex and resource-intensive, with substantial costs, time, and expertise required for data collection, annotation, and curation \cite{sambasivan2021everyone}.  Increasingly, \emph{synthetic data}---artificial data generated using models or algorithmic techniques~\cite{pdpc2024policy}---is seen as an appealing alternative~\cite{nikolenko2021synthetic}.  OpenAI~\cite{achiam2023gpt}, Apple~\cite{apple, gunter2024apple}, Microsoft~\cite{abdin2024phi}, Google~\cite{team2023gemini}, Meta~\cite{llamameta}, and IBM~\cite{IBM} have all reported using synthetic data in their AI development pipelines or advertised the ability of their models to create synthetic data. \looseness=-1

Synthetic data is not new. Explorations into data simulation to address data scarcity date back at least to the 1970s~\cite{nikolenko2021synthetic}. The idea began to gain momentum with the introduction of practical methods for data generation, like generative adversarial networks \cite{goodfellow2020generative} and variational autoencoders \cite{kingma2013auto}. Throughout much of the past decade,
the use of synthetic data was viewed as a way to mitigate concerns about fairness, bias, and privacy. Proponents argued
that privacy concerns could be addressed by replacing traditional datasets with differentially private, synthetic alternatives~\cite{abay2019privacy, xie2018differentially, dahmen2019synsys, jordon2018pate} and that model biases could be mitigated by augmenting traditional datasets with simulations of data from underrepresented groups or rare scenarios \cite{shin2018medical, jaipuria2020deflating}. The latter approach was especially popular for applications like facial recognition, where collecting diverse data can be prohibitively challenging \cite{Kortylewski2019analyzing,bae2023digiface1m}. Practitioners have since applied synthetic data in applications within the robotics \cite{huang2022inner, liang2023code}, automotive \cite{yang2020surfelgan}, finance \cite{assefa2020generating}, and healthcare~\cite{synthea, meeker2022case} industries, among others. \looseness=-1

Now, the widespread availability of large, off-the-shelf generative models has given practitioners the ability to automate the creation of synthetic data without requiring extensive domain-specific expertise or custom-built tools. This has made synthetic data more accessible and rapidly expanded the scope of its use in AI development~\cite{jordon2022synthetic,liu_best_2024}.  It is increasingly common for researchers and practitioners to use one large generative model (which we refer to as an \emph{auxiliary model} throughout this paper) to generate synthetic data that is used to train or evaluate another (the \emph{primary model}).  At the same time, the nature of data is evolving.  Auxiliary models are used not only to expand training datasets~\cite{luo2023wizardmath}, create benchmarks~\cite{hartvigsen_toxigen_2022}, and develop test cases~\cite{feng2023factkb, perez_red_2022}, but also to generate scores or labels for model outputs~\cite{zheng2023judging, dubois2024alpacafarm}, a task previously performed by human annotators or with simple automated approaches. And since auxiliary models can dynamically generate responses to a primary model's outputs in real-time, synthetic data no longer needs to be static but can be generated interactively as needed. \looseness=-1

Despite concerns that have been raised over the risks of synthetic data~\cite{whitney_real_2024, agnew2024illusion}, policy guidance and best practices for its responsible use have not kept pace with shifting trends. Much of the policy guidance is still centered on privacy, with more widespread applications of model-generated data just starting to receive attention.  For instance, Singapore's Proposed Guide to Synthetic Data, released
in July 2024, focuses on the benefits and risks of synthetic data as a privacy-enhancing technology~\cite{pdpc2024policy}.  The EU's General-Purpose AI Code of Practice touches only briefly on synthetic data, with a recommendation added in the November 2024 draft to document ``a description of the methods, if any, used to synthetically generate training data'' and ``the name(s) of any AI model(s) or system(s) used to synthetically generate training data'' \cite{euaicodeofpractice}. \looseness=-1

One challenge in creating comprehensive guidance on synthetic data is a lack of understanding of practitioners' current workflows.  With the landscape rapidly evolving and the use cases so varied, we do not have a clear picture of the full spectrum of practitioners' motivations for using synthetic data, the properties they hope it will satisfy, or how (or even if) they validate the data they create. To address this gap, we explore the following research questions: \looseness=-1

\begin{itemize}
    \item \textbf{RQ1:} What are practitioners’ motivations for using synthetic data in the AI development pipeline?
    \item \textbf{RQ2:} What are practitioners’ current practices, desiderata, and challenges when generating synthetic data?
    \item \textbf{RQ3:} What are practitioners’ current practices and challenges when validating synthetic data?
    \item \textbf{RQ4:} What limitations and ethical considerations arise with the use of synthetic data?
\end{itemize}

We draw on a two-phase interview study.  We first conducted semi-structured interviews with 19 practitioners actively engaged in the development of generative AI models or systems to understand how they use synthetic data across the development pipeline. To further explore the limitations and ethical considerations of the practices identified, we conducted an additional 10 interviews with subject matter experts experienced in responsible AI aspects of dataset creation or model evaluation, scaffolding the discussion with three vignettes based on use cases observed in phase 1.

While synthetic data is a broad concept, in order to focus on new and evolving practices, we scope our study to the use of data (including labels and scores of outputs) produced by auxiliary generative models for use in the AI development pipeline.  While we use terms like `real' or `human-generated' data for distinction throughout this paper, we acknowledge that synthetic data is also shaped by human design and intervention \cite{lee2025ontological, jacobsen_machine_2023} and that, in practice, there is a spectrum between `real' and `synthetic' data. \looseness=-1

Our interviews reveal how auxiliary models are now embedded in nearly all stages of the AI development pipeline, with synthetic data increasingly substituted for traditional human-generated data and labels.
Practitioners perceive synthetic data as a promising and indispensable tool to address data scarcity challenges, resource constraints, and high data collection costs across a range of critical tasks, including evaluating harms, diversifying datasets to reduce bias, preserving privacy when working with sensitive data, and scoring model outputs.
However, practitioners faced challenges in understanding and controlling the outputs of auxiliary models and generating data that accurately depicts members of underrepresented groups.  Although practitioners acknowledged the importance of data validation, they struggled to define what makes synthetic data `good' and reported that most validation currently takes the form of `spot-checking' or `eyeballing.'  Responsible AI experts noted additional limitations, including risks of stereotyping and the removal of avenues for data subjects to exercise agency over their data. 
We reflect on these findings and their implications for the evolution of data practices in section \ref{sec:discussion} and conclude by highlighting opportunities for future FAccT research to support the responsible use of synthetic data.  

Our paper makes three main contributions:
\begin{itemize}
    \item We provide an empirical account of the ubiquitous and inconsistent integration of synthetic data into modern AI development pipelines.
    \item We highlight limitations and ethical considerations of synthetic data use and propose avenues to study its impact.
    \item We provide considerations for the responsible use of synthetic data aimed at both practitioners and policymakers.
\end{itemize}

\section{Related Work}

In this section, we provide an overview of the uses of synthetic data in modern AI development pipelines, draw attention to critiques of synthetic data, and engage with scholarship examining industry practices around AI development and data production.  \looseness=-1

\subsection{Synthetic Data Usage in Modern AI Development}
\label{sec:rw-syn-data-gen}

Advances in generative AI have led to a rapid expansion of the use of synthetic data. We briefly review use cases from the literature and direct readers to the surveys of Jordon et al. \cite{jordon2022synthetic} and Liu et al. \cite{liu_best_2024} for a comprehensive overview. \looseness=-1

Within the training stage of the AI pipeline, synthetic data is used to improve generative model capabilities. For instance, researchers have designed synthetic datasets geared to teaching models mathematical reasoning~\cite{luo2023wizardmath,yu2023metamath} and computer programming~\cite{hu2023instructcoder}.  Synthetic language-image datasets have been designed to improve the performance of multi-modal applications \cite{liu2024visual}, while datasets of synthetic multilingual question-answer pairs have been used to improve cross-lingual performance \cite{riabi2020synthetic, shakeri2020towards}. Synthetic data is also commonly used for knowledge distillation, which aims to transfer knowledge from a large model to a smaller one~\cite{Hinton2015DistillingTK,gou2021knowledge}.  Across these varied use cases, synthetic data is positioned as a way to augment real-world datasets and introduce controlled variability \cite{liu_best_2024}. \looseness=-1

Within the evaluation stage, researchers have generated synthetic test cases to evaluate models for characteristics like factual consistency \cite{feng2023factkb} and harmful behavior \cite{perez_red_2022}, and have published synthetic benchmark datasets, for instance, for detecting harmful output~\cite{hartvigsen_toxigen_2022}. Synthetic test cases can take the form of static model prompts or, since auxiliary models can generate responses on the fly, interactive simulations of a user's interactions with a model or system. Beyond creating test cases, auxiliary models are increasingly used to evaluate or score outputs from a primary model~\cite{gilardi2023chatgpt,dubois2024alpacafarm}, an idea colloquially known as `LLM-as-a-judge' \cite{chiang2023can}. We view this as a form of synthetic data as the scores produced by the auxiliary model take the place of labels that often would have been collected from human evaluators.

\subsection{Critical Studies of Synthetic Data}
\label{sec:rw-critical-data-studies}

Scholars have challenged the assumption that data offers an unmediated reflection of reality, revealing instead its deeply constructed nature. As Gitelman~\cite{gitelman2013raw} argued, data is never `raw.'

The abstraction introduced by synthetic data creates further distance from the material realities seemingly represented in datasets, raising the question of whether and to what extent synthetic data is a reflection of reality.  \looseness=-1

Researchers have invited a critical examination of how synthetic data may introduce new risks and exacerbate existing ones in ways that may evade scrutiny \cite{jacobsen_machine_2023}.
Susser and Seeman \cite{susser_dialogue_nodate} caution against positioning synthetic data as free from ethical scrutiny while enabling its role in accelerating unchecked AI development.
Helm et al. \cite{helm_generating_2024} argue that synthetic data operates as a discursive device, legitimizing a shift from data collection to data generation without addressing the ethical and epistemological harms intrinsic to the models that produce this data.
Synthetic data’s ostensible detachment from real-world individuals can mitigate privacy concerns, but it simultaneously forecloses meaningful engagement with the people and communities most affected by AI systems \cite{susser_synthetic_2024}.
By severing relationships between data subjects and the systems that act upon them, synthetic data erodes pathways for accountability, effectively shielding its use from critical oversight \cite{fitzgerald_why_2024}
and adding layers of opacity that obscure how synthetic data influences downstream decision-making. 
Furthermore, the promise of synthetic data as a fix-all solution can undermine democratic approaches to data governance, silencing opportunities for public participation and reflexive deliberation \cite{whitney_real_2024}.
To interrogate these risks and weigh them against potential benefits, there is a need to examine both the conditions of data production and the auxiliary models that are used to generate this data \cite{wiehn_synthetic_2024}.


\subsection{AI Development Practices}

\label{sec:rw-data-practices}


The logics, methods, and techniques that define the field of AI are inseparable from its values, social norms, and organizational imperatives~\cite{elish2018situating}. Decisions made throughout the AI development pipeline are embedded in broader social, cultural, and organizational contexts~\cite{vertesi2011value}. To identify points or intervention or improvements, it is crucial to understand the context and challenges of AI development. A growing line of research within the FAccT, CSCW, and broader HCI communities aims to shed light on the practices, challenges, and needs of practitioners developing AI systems. This line of work has explored practices around responsible technology development~\cite{holstein2019improving, deng2022exploring, deng2023investigating, rakova2021responsible,passi2018trust,madaio2020co,widder2023s,yildirim2023investigating,madaio2022assessing} and examined data practices specifically~\cite{wongsuphasawat2019goals, han2023data, crisan2020passing, ruddle2023tasks, muller2019data, zhang2020data, kapania_hunt_2023, heger2022understanding}, exposing the challenges faced by practitioners throughout data curation \cite{han2023data}, exploratory data analysis \cite{wongsuphasawat2019goals}, annotator selection \cite{kapania_hunt_2023}, and data documentation~\cite{heger2022understanding}.  The role of organizational factors 
has been a common theme~\cite{madaio2020co, madaio2022assessing, deng2023investigating, widder2023dislocated}.  Researchers have emphasized the discretionary choices that shape data work, such as how tasks are formulated, what data is collected and annotated, how data quality is measured, which errors are acceptable, and what is communicated to stakeholders \cite{passi_data_2017}.  Muller et al. \cite{muller2019data} examine how trade-offs and compromises are made when intervening with data, while Sambasivan et al. \cite{sambasivan2021everyone} report how contextual factors and incentives for prioritizing model development over careful data development result in downstream harms  \cite{koesten_everything_2020}. \looseness=-1

In concurrent research, Qian et al. \cite{qian2024evolution} studied practices within Google around the adoption of large language models for data curation.
They find that data workers emphasize the importance of high-quality data, but have trouble defining data quality for generative outputs without clear ground-truth references. They describe an emerging dataset hierarchy where the practice is shifting from the usual human-labeled `golden datasets' to model-generated `silver datasets' and `super-golden datasets' that are created and rigorously evaluated by product managers, policy makers, engineers, and other experts.  Our study complements this work by exploring a wider range of use cases for synthetic data across multiple organizations to understand practitioners' motivations for using synthetic data; their current practices, desiderata, and challenges when generating and evaluating synthetic data; and limitations and ethical considerations that arise. \looseness=-1

\section{Methods}
\label{sec:methods}

To investigate our research questions, we developed a two-phase interview study, conducting semi-structured interviews with a total of 29 participants from 14 organizations across the United States. The first phase focused on understanding AI practitioners' dataset, model, and system development and evaluation practices for generative AI. We asked participants to describe their AI pipeline for a particular project, including the types and sources of data they used. We then asked specifically about their use of synthetic data and labels, investigating their motivations, how they generate the data, the validation processes they follow, the challenges they encounter, and their perceptions of any limitations. \looseness=-1

To further explore the limitations and ethical considerations of the practices identified as well as paths forward, the second phase of interviews focused on understanding responsible AI (RAI) experts' perspectives on the sociotechnical dimensions of synthetic data use. After asking about participants’ areas of focus within RAI, we asked them to reflect on the ways synthetic data intersects with privacy, fairness, consent, and other RAI concerns, discussing whether synthetic data could help mitigate existing challenges and whether it introduces new risks and trade-offs. To scaffold our discussions, we distilled the use cases of synthetic data mentioned in phase 1 interviews into three hypothetical scenarios: using synthetic data to fine-tune models, create evaluation datasets, or simulate user interactions.
For each scenario, participants reflected on the practical and ethical implications of the practice, including any concerns and trade-offs involved in adopting synthetic data over traditional methods. Refer to Appendix \ref{sec:appendix} for study materials, including the vignettes. \looseness=-1

All interviews were conducted over video conferencing software in English between May and August 2024. Each session lasted 40--60 minutes, and participants were compensated with a gift card of \$75 USD. We recorded field notes and video recordings, which were transcribed for analysis. See Section~\ref{sec:ethics} for a discussion of research ethics considerations. \looseness=-1

\textbf{Recruitment.} We recruited participants until we reached saturation through advertisements on social networks such as X (formerly Twitter) and LinkedIn, direct emails to contacts in our professional networks, and snowball sampling. For phase 1, we recruited 19 participants from industry and academia who were actively engaged in the development of generative AI models or products. We refer to these participants as \textit{practitioners}. Most practitioners focused on product development and research, in roles such as applied scientist, software engineer, linguist, or manager. 
For phase 2, we recruited 10 participants who were focused on RAI research or development and had experience in RAI aspects of dataset creation or model evaluation, whom we refer to in this paper as \textit{RAI experts}.  These participants worked in technology companies, non-profits, and academia. Table~\ref{tab:participant_info} includes more details on participants. 

\begin{table*}[t]
\footnotesize
  \centering
   \begin{tabular}{p{2.8cm}p{3.8cm}|p{2.7cm}p{3.8cm}} 
     \multicolumn{2}{c|}{\textbf{Practitioners (\textit{n} = 19)}} & \multicolumn{2}{|c}{\textbf{RAI Experts (\textit{n} = 10)}} \\ \midrule
    Industry (\textit{n = 15}) \newline
    Academia (\textit{n = 4})
    &
    
    Researcher (\textit{n} = 7) \newline
    Applied Scientist (\textit{n} = 4) \newline
    ML Scientist (\textit{n} = 2) \newline
    Product Manager (\textit{n} = 2) \newline
    Linguist (\textit{n} = 1) \newline
    Red-teaming Lead (\textit{n} = 1) \newline
    Research Manager (\textit{n} = 1) \newline
    Software Engineer (\textit{n} = 1)
    &

    Industry (\textit{n = 3}) \newline
    Academia (\textit{n = 6}) \newline
    Non-profit (\textit{n = 1})
    
    & 
    Researcher (\textit{n} = 4) \newline
    Professor (\textit{n} = 3) \newline
    Data Operations Lead (\textit{n} = 1) \newline
    Policy Researcher (\textit{n} = 1) \newline
    Program Manager (\textit{n} = 1)
  \end{tabular}
  \vspace{1em}
  \caption{Summary of participants' roles. Participants spanned a total of 14 organizations across the United States. In reporting our findings, we use identifiers beginning with `P' for practitioners and `E' for responsible AI experts.}
  \vspace{-2em}
  ~\label{tab:participant_info}
\end{table*}

\textbf{Analysis.} 
We followed a reflexive thematic analysis approach inspired by Braun and Clarke \cite{braun2012thematic, braun2019reflecting}. Our analysis was an inductive and iterative process guided by our research questions. The first author read each interview transcript multiple times to get familiar with the data, then extracted codes that aligned with the concepts in our research questions. The entire research team met regularly to discuss ambiguities and to define themes based on our initial codes. We surfaced 762 first-level codes. As we generated themes from the codes, we also identified categories with a description and examples of each category. In the early stages of our codes-to-themes process, we generated nine domain categories (stages of use, motivations, objectives, synthetic data generation techniques, validating synthetic data, challenges, limitations, trade-offs, and assumptions). These categories were also iteratively refined through meeting, diverging, and synthesizing into four top-level categories organized by our research questions and presented in our findings. \looseness=-1

\textbf{Limitations.} Because of the sensitive nature of our study and the current competitive environment in AI, recruiting practitioners to discuss their data practices was challenging. We were limited in the number of participants we could recruit, the breadth of organizations they represented, and the global diversity of those participants.  In particular, all practitioners were based in the United States. Additionally, most practitioners in our study focused on language-based technologies and, therefore, text data. The three exceptions to this focused on multimodal data (text and images), image data, and biological sequence data, respectively.  These factors potentially skewed the data practices we observed and challenges highlighted in our findings.  Additionally, while part of our original motivation for recruiting RAI experts was to aggregate guidance for synthetic data use, we found that concrete best practices have yet to emerge. We therefore pivoted this discussion to steps the community can take to develop best practices in light of our findings. \looseness=-1

\section{Findings}

We first examine how practitioners use auxiliary generative models to produce synthetic data and their motivations for doing so (RQ1).  We next describe their current practices, desiderata, and challenges when generating (RQ2) and validating (RQ3) synthetic data.  Finally, we discuss general limitations of and ethical considerations for synthetic data (RQ4). \looseness=-1

\subsection{The Expanding Role of Synthetic Data}
\label{sec:role}

Our interviews reveal a reconfiguration of AI development workflows, with auxiliary models now embedded in nearly all stages of the AI development pipeline and synthetic data increasingly substituted for traditional, human-generated data and labels. Practitioners articulated a wide range of motivations for incorporating synthetic data, including its potential ability to scale, perceived performance improvements, and increased controllability. \looseness=-1

\subsubsection{Use Cases Across the AI Pipeline.}
\label{sec:usecases}

Practitioners described a breadth of use cases for synthetic data, closely aligned with those proposed in the literature (Section~\ref{sec:rw-syn-data-gen}).
Within the training stage of the pipeline, practitioners frequently used auxiliary models to produce large datasets or augment existing datasets for pre-training new primary generative models, particularly in domains where data scarcity or sensitivity limited access to training datasets. Participants explained how synthetic data enabled them to create specialized corpora, such as synthetic legal documents or medical records, which would otherwise remain inaccessible due to privacy and/or compliance requirements. Synthetic data was also used to fine-tune existing primary models for specialized tasks. P15, for example, generated instruction-following datasets to align model outputs with safety requirements, while P12 used a larger auxiliary model as an `oracle' to perform knowledge distillation ~\cite[cf.][]{gou2021knowledge} to train a smaller, more efficient model.

In the evaluation stage, practitioners turned to synthetic data to assess both the performance and safety of the primary model. Participants described how auxiliary models were used to generate diverse test cases to evaluate the primary model's ability to handle a range of input types, including cases that might not be well-represented in real-world datasets. For emerging tasks that lacked established datasets or evaluation metrics, synthetic data promised a solution for creating custom benchmarks. Sometimes, synthetic evaluation data consisted of fixed queries, while sometimes, practitioners used auxiliary models for simulating user interactions dynamically, such as sequences of queries or dialogues. Such simulated interactions were integrated into automated red-teaming efforts in which auxiliary models generated adversarial inputs designed to test the primary model's adherence to specific safety principles. \looseness=-1

Beyond generating inputs for evaluation, auxiliary models were also increasingly used to evaluate the quality of the primary model's outputs by producing ratings or scores (as in LLM-as-a-judge ~\cite{zheng2023judging}), a task that was previously typically carried out by human annotators or more conventional automated metrics (e.g., \cite{papineni2002bleu, fabbri2021summeval}). For example, P11 was using an auxiliary model for red-teaming purposes: specifically, to evaluate whether the primary model was producing harmful outputs. They explained how the model \textit{``output obviously needs to be categorized, and that is done using [auxiliary] models. So we have kind of like scorers [...] across different harm categories that sort the outputs, which I think could count as synthetic data insofar as it's making a judgment on the output of the [primary model].''}  
In some cases, the scoring prompts and evaluation rubrics themselves were developed using an auxiliary model.  Initially, several participants did not consider their use of auxiliary models for scoring model outputs as a form of synthetic data (or `data' at all), but on further reflection, they came to identify these labels as `synthetic evaluation data.' \looseness=-1

\subsubsection{Perceived Promises of Synthetic Data.}

Practitioners brought up a variety of reasons why they expected synthetic data to be beneficial, which we describe here. However, as we explore in Sections~\ref{sec:genchallenges} and~\ref{sec:valchallenges} when reflecting on the challenges of generating and validating synthetic data, these promised benefits were not always realized in practice.

In describing motivations for their use of synthetic data, practitioners emphasized that auxiliary models enabled them to overcome data and resource constraints and thus significantly accelerate their development and evaluation processes. \looseness=-1

The relatively lower cost and time requirements of using synthetic data emerged as key factors driving its appeal. For example, P15 described how synthetic data allowed them to produce large volumes of expert-level math or science training data without needing to recruit mathematicians or scientists. Participants emphasized that model training is better resourced than model evaluation, making synthetic data an especially attractive option for evaluation purposes. P3, who specialized in developing an automated red-teaming pipeline, provided a compelling comparison: Manual red-teaming often required hiring cybersecurity experts, each spending up to forty-five minutes interacting with the system to generate a single sample. In contrast, automated red-teaming incurred substantially lower costs---primarily initial setup and model inference expenses. The cost differential made a ``\textit{cheap and dirty}'' automated approach far more attractive.  \looseness=-1

The use of human annotators, meanwhile, was limited by the growing needs across teams, especially for evaluation.  P5 and P11 noted that tight timelines and organizational demands required generating thousands of prompts and testing models within just a few weeks, with dozens of teams in P5's organization conducting weekly model evaluations.  They noted that scoring large volumes of model outputs weekly was infeasible with human reviewers. \looseness=-1

Participants also embraced synthetic data for its potential to boost performance and provide a competitive edge. P17's decision to adopt synthetic data was inspired by another team within their company that had successfully used synthetic data for evaluation. As they explained: ``\textit{We had a pretty good idea going into our own testing that this type of prompt was appropriate, or was going to be fairly accurate, hopefully, if we did things correctly.}'' The anticipated performance gains also helped some participants justify its use to upper management. Several participants emphasized the competitive advantage of synthetic data offered in the tech industry, particularly for pre-training new, primary foundation models. P18 underscored how synthetic data would facilitate differentiation in model development, even if the approaches for generating it were publicly known:
    ``\emph{Even if I tell you exactly how I generated my synthetic data, I will bet all my money that there's no way you can generate the same thing... we're not going to have the same model. And so it's an easy way to differentiate between companies and models, especially when we all start with the same boat, which is web data.}'' \looseness=-1

Practitioners were particularly drawn to the parameterized and (ostensibly) controlled nature of synthetic data generation, which was perceived as a stark contrast to the unpredictability of real-world data collection and labeling processes.  For example, P1, who was developing a benchmark for hate speech and toxicity detection, noted that publicly available datasets frequently exhibited skewed representations, favoring particular identity groups or containing disproportionately high levels of toxic content compared to non-toxic examples. These imbalances limited practitioners' ability to effectively evaluate their primary models. Synthetic data, then, provided a promising avenue to address these shortcomings by enabling more `controlled' data generation.  As P18 put it, with synthetic data, ``\textit{you have full control of what you're generating.}'' However, as we describe in section \ref{sec:generating}, `full control' was not always achievable in practice.

Practitioners also viewed synthetic data as a way to simplify compliance requirements related to data protection. P15 explained how generating synthetic data allowed them to avoid the ``\textit{lengthier processes}'' involved in securing data rights. Similarly, P4 noted that ``\textit{feedback from real customers has to be deleted after 30 days and is subject to GDPR, data subject requests, and things like that, but the synthetic version does not have those same encumbrances.}'' While highlighting this as an advantage, practitioners acknowledged the ``\textit{gray area}'' with this approach, noting that \textit{``legal issues''} with regards to synthetic datasets are far from resolved.

\subsection{Generating Synthetic Data}
\label{sec:generating}

We next explore current practices, desiderata, and challenges when \emph{generating} synthetic data (RQ2).

\subsubsection{Desiderata for Synthetic Data.}

The two desiderata most commonly identified by practitioners were diversity and resemblance to `natural' or `real' data. Diversity was often described in terms of capturing a range of scenarios, input types, or demographic characteristics, which participants considered important to enable primary models to generalize effectively across varied real-world contexts. 
Natural-looking data was described as closely resembling human-generated content, but without replicating the training data of the auxiliary model. For instance, in the context of generating data for automated red-teaming, P3 explained that the average user would not make an explicitly harmful query like ``\emph{how to make a bomb}.'' Instead, more frequent ``\textit{naturally occurring queries}'' might trigger more subtle forms of harm relating to, for instance, interpersonal bias in the workplace.

\subsubsection{Approaches to Data Generation.}

Practitioners chose different approaches to synthetic data generation, with distinct trade-offs between control and flexibility. Some preferred a constrained approach that limited the scope of generation to maintain more oversight over outputs. Most commonly, this involved having the auxiliary model generate variations (\textit{e.g.,} paraphrases) of manually curated seed examples.  P13, for example, generated synonyms for a list of successful red-teaming prompts, arguing that this structured approach supported the rigor required for controlled experimentation by allowing them to monitor how language shifts impacted the primary model's responses.  Another common constrained approach was to design structured templates that dictate the format the auxiliary model's output should take, for instance, steering the model output toward specific topics or demographic groups.

Other practitioners were willing to sacrifice some control for the flexibility afforded by more open-ended approaches. One such approach was to generate synthetic data by having the auxiliary model take on a simulated persona reflecting specific demographics or scenarios. 
Practitioners reasoned that this approach allowed for exploration across a broader set of possible outputs, offering more diversity and resemblance to natural data. While there was a risk of introducing lower-quality irrelevant data, practitioners adopting open-ended approaches would iteratively rewrite the prompt for the auxiliary model if the generated data did not match their expectations.

Different techniques were needed to generate the (synthetic) scores or ratings used in evaluation. The most common approach was to specify annotation guidelines for the auxiliary model to follow, similar to those that would be given to human raters. P10 believed this approach is slowly emerging as the gold standard for evaluating model-generated summaries: ``\textit{It’s not perfect, but for better or worse, it’s the [method] most calibrated with human judgments that exists. You give instructions on how GPT-4 should evaluate the summary, present the source document and the generated summary, and then GPT-4 assigns a score from 1 to 7 based on the rubric provided.}'' \looseness=-1

Participants described an ad-hoc approach to selecting which auxiliary model to use for data generation, with choices driven by immediate needs and availability rather than systematic criteria. Many relied on models readily accessible within their organizations or those considered ``\textit{state-of-the-art.}'' P9 explained ``\textit{we've always operated on the idea that the newest model is going to be the best and the most helpful for our customers.}'' P4 described choosing ``\textit{the most powerful LLM that [their team] feels is economically reasonable to prompt.}'' Teams often matched models to task complexity, as P18 noted, using GPT-3.5 for ``\textit{straightforward tasks, such as cleaning up typos or translating text},'' while reserving GPT-4 quotas for more complex activities, such as generating math problems. 

Typical large, off-the-shelf models undergo alignment processes intended to ensure helpful and safe responses~\cite[cf.][]{ouyang2022training}. By design, this would limit the model's capacity to generate responses containing stereotypes or potentially offensive content.  
P5 recalled trying to use an `aligned' auxiliary model for safety testing.  When the primary model outputted ``\textit{I’m sorry, I cannot generate stereotypes,}'' the auxiliary model would affirm with responses like ``\textit{Great, you should not,}'' derailing the evaluation. 
For this reason, when working on safety-related tasks, such as fine-tuning a primary model to generate safe responses or evaluating for harms, participants often used what they believed were `unaligned' auxiliary models. \looseness=-1

\subsubsection{Challenges with Synthetic Data Generation.}
\label{sec:genchallenges}

Participants identified several challenges in generating high-quality synthetic data. Most could be traced to the limitations of the auxiliary models themselves since, as P16 emphasized, synthetic data inevitably reflects the capabilities and limitations of the auxiliary model. Participants described auxiliary models as brittle and highly sensitive to minor changes in prompts. P6 referred to prompt engineering as \textit{"more of an art than a science,"} noting that small adjustments (e.g., reordering words, changing punctuation, or rephrasing sentences) would significantly impact quality. Practitioners found themselves engaged in constant experimentation, attempting to (prompt) ``\textit{engineer away}'' odd behaviors or inconsistencies. 

While natural-looking data was desirable, some practitioners questioned whether synthetic data distributions could closely resemble real data. Many were skeptical,
and P7 cautioned that ``\textit{one would hope that the distributions are quite similar, but they're not going to be exactly the same on the real and synthetic data.}''

While controllability was viewed as a major benefit of synthetic data, practitioners often struggled to understand or control the outputs of the auxiliary models. P12, who used an auxiliary model for model distillation, acknowledged how its opacity left their team uncertain about whether the primary model was learning the correct function, noting that ``\emph{it is hard to say exactly what’s going on here.}''
P5 encountered a similar issue when using an auxiliary model to create socio-cultural identity prompts to seed safety testing for their primary model. They intended the simulator to generate responses reflecting specific stereotypes for testing purposes but discovered that ``\emph{all the responses were about tacos.}'' Upon manual inspection, their debugging revealed how ``\textit{it was an issue in the pipeline where a single few-shot example that happened to mention Mexican food and tacos [were] being injected and overriding some other stuff.}''  And P13, the practitioner who had generated synonyms for a list of successful red-teaming prompts, acknowledged that even with this targeted method, they couldn’t fully control the model’s semantic interpretations: 
    ``\emph{If you just think of the word `red,' in addition to giving us colors, [the auxiliary model] gave us this more conceptual kind of bloody. It could’ve gone in a political direction with that. It could’ve gone in a racist direction with that... That might be in the [data], and we currently don't know.}''
    \looseness=-1

Finally, while synthetic data was thought to have the potential to address gaps in representation by generating data for underrepresented groups, many practitioners described how synthetic data failed to accurately depict smaller populations.
P8 noted that this increased the risk of privacy violations by producing outputs that could potentially identify individuals within these groups. For this reason, E20 was apprehensive about any use of synthetic data in the evaluation stage; if synthetic test cases do not adequately reflect minority groups, the resulting metrics may provide a misleading sense of confidence in the primary model's safety or performance.

\subsection{Validating Synthetic Data}
\label{sec:validating}

We now explore practitioners’ current practices and challenges when \emph{validating} synthetic data (RQ3).

\subsubsection{Approaches to Data Validation.}

Practitioners described the validation of synthetic data as ``\emph{subtle work}'' (P18) that was subjective and required them to rely on their intuition.
Manual inspection---also referred to as ``\emph{spot-checking}'' or ``\emph{eyeballing}'' by participants---emerged as the most common validation method.
P3 described a periodic monitoring routine in which they inspected examples every few weeks to ``\textit{make sure the data is as it should be, and that things are going smoothly.}'' During these inspections, they often discovered cases where the model ``\textit{was not following the prompt correctly and things don't make sense.}'' Similarly, P5, who focused on measuring harms, shared their process of manually reviewing datasets to check for alignment with instructions, explaining ``\emph{I would generate a dataset and then scroll through it, asking, `Does this make sense? Is it following the instructions we’ve given it?'}'' Practitioners acknowledged that this approach made it challenging to maintain consistency, especially when the data was inspected by different team members. \looseness=-1

Alternative methods beyond manual inspection were rarer.  P4, whose work involved simulating user data, described how they ``\textit{don't yet have a methodology for evaluation}'' of synthetic data. Instead, they focused on the usefulness of the data in downstream applications.  A few, including P15 and P18, conducted small-scale studies with human experts (e.g., doctors or trust and safety professionals) to validate synthetic data for specialized tasks. These participants also performed ablation experiments by generating small synthetic datasets and evaluating their impact on the primary model’s performance when used for training. P7 described how their team used the internal confidence score of the auxiliary model in validating its own output, noting the circular logic. They set thresholds to classify outputs, treating synthetic data with confidence scores above 90\% as high quality and those around 30\% as low quality. \looseness=-1

\subsubsection{Challenges in Data Validation.}
\label{sec:valchallenges}

Participants described validation of synthetic data as a consistent bottleneck. Ironically, the qualities that made synthetic data attractive--such as its scale, diversity, and the ability to simulate data for rare scenarios--also made it exceptionally challenging to validate since, as described above, validation was typically manual. \looseness=-1

Certain types of data were especially challenging to inspect manually. P8, who worked in medical imaging, described the challenges of confirming whether synthetic images accurately reflected specific medical conditions, particularly those with subtle visual markers.
They collaborated with a resident doctor to assess the realism of their data. Practitioners working in high-stakes domains, like harmful content detection or red-teaming, encountered similar challenges. \looseness=-1

Another significant challenge was defining what qualifies as `good' synthetic data, especially in contexts where no real data exists to guide conceptualization. As described above, practitioners sought data that appeared diverse and natural-looking. Yet, as participants noted, it was difficult to articulate what these criteria mean. 
For example, P5, who focused on establishing safety evaluations for generative AI products, described how their workflow lacked user data to help gauge the diversity and distribution of model outputs.
Practitioners often defaulted to defining diversity as the generation of non-redundant examples, such as P16, who described how
``\emph{the main way [they] defined diversity was how many unique examples could be generated. If [they] call the generator model a thousand times, how many redundant examples were generated?}''
However, simply avoiding redundancy did not guarantee that the data covered a meaningful range of variation.
P3, who focused on detecting stereotypes in primary model outputs, found it challenging to define the types of stereotypes they aimed to capture and operationalize these criteria within prompts.
And P17 struggled to validate whether the synthetic data accurately reflected underrepresented groups, noting the lack of clear indicators to confirm that the auxiliary model was following instructions or capturing the intended range of diversity.

Lastly, practitioners highlighted that their companies prioritized scaling data generation to keep pace with model development demands, relegating validation to a lower priority. P3 expressed this tension: ``\textit{Honestly, people just don’t want to do [the validation] right now because of the urgency associated with everything.}'' While this participant recognized the risks of insufficient validation, they felt pressured to move forward to meet internal deadlines, hoping to address data quality issues later if time and resources allowed.  \looseness=-1

\subsection{General Limitations and Ethical Considerations of Synthetic Data}
\label{sec:limitations}

Beyond the challenges specific to generating and validating synthetic data, practitioners and RAI experts identified several broader limitations and ethical considerations with the use of synthetic data. \looseness=-1

\subsubsection{`Chaining' Auxiliary Models.}

Practitioners described using the same or similar auxiliary models to produce both training and evaluation data, or to generate evaluation data and score the primary model's outputs. This practice of `chaining' auxiliary models raised concerns about the risk of amplifying biases at different stages of the AI development pipeline.  P10 gave a hypothetical example where the auxiliary model was the same as the model being evaluated: ``\emph{GPT-4 as a scorer and asking it to evaluate a GPT-4 based prediction and comparing that to a LLaMA prediction. There's a high chance GPT-4-based prediction will be scored higher because your scoring module is also GPT-4.}''  E24 reinforced how the cumulative effect of seemingly \textit{``minor problems or morally neutral outcomes''} could become significant when repeated on a massive scale: ``\textit{You've done something that maybe is not that bad, but you have done it 10,000 times instead of 100 times, and you have also done it in a way where you are not really sure what the gaps are and what the dangers are.}'' They argued that chaining models creates feedback loops that mask issues that would emerge through more diverse evaluation methods. \looseness=-1

Using generative models across multiple stages also introduced the risk of creating distribution shifts that are not easily identifiable without real data to compare against. Both practitioners and RAI experts expressed concerns about the long-term risk of `model collapse'~\cite[cf.][]{Shumailov2023modelcollapse} 
where the repeated use of synthetic data at the training stage causes models to deviate from real-world data distributions, leading to cascading errors and degraded performance over time. Opinions on this risk varied: some participants believed that combining high-quality synthetic data with real data could sustain model development without severe degradation, while others viewed this distribution shift as a slippery slope that resulted in compounding errors that threatened the integrity of models over time. The lack of consensus around this topic was described as a challenge to both understanding the problem and formulating appropriate solutions. \looseness=-1

\subsubsection{Removed Avenues for Exercising Agency.}

RAI experts brought up that the use of synthetic data undermines stakeholders' ability to exert meaningful agency over AI systems. They emphasized that synthetic data creates distance between individuals and the data that was originally derived from their activities, decisions, or behaviors. This layer of abstraction erodes opportunities to contest the ways that datasets are used. As E24 noted, ``\textit{the link between a person and their data and a training dataset}'' is arguably one of the most effective opportunities for participation or redress. \looseness=-1

Several RAI experts remarked on the extractive nature of generating synthetic data using auxiliary models that may have been developed through the expropriation of the work of data subjects. E27 linked these concerns to ongoing debates in the digital humanities, where artists, writers, and creators have called for ethical engagement with data derived from their work, critiquing practices that commodify their intellectual and creative output~\cite[cf.][]{jiang2023ai}. E28 suggested that the purported privacy-preserving benefits of synthetic data often come at the expense of others, explaining that for auxiliary models to generate data about a specific group, they require relevant information in the training data: \textit{``so how good this [data is] is a matter of how well it is violating someone else's privacy, right?''}  \looseness=-1

\subsubsection{Stereotyping and Cultural Hegemony.}

Participants highlighted
risks of using identity-related prompting (i.e., including a series of adjectives to describe a simulated user's identity) when generating data.
E25 noted that this practice could cause an auxiliary model to either produce stereotypical patterns of behavior or generate overly generic responses due to its alignment process.
P6 and E26 both noted that synthetic data intended to reflect identity often lacked nuance and depth, in part because auxiliary models rely on data \emph{about} a group rather than data generated \emph{by} members of the group~\cite[cf.][]{Wang2024llms}. As P6 put it:
    ``\emph{Most of that data is in a tone or a stance, or a perspective of being about these different groups of people, and not by these groups of people, and so there's a lot of cases where we want to generate data as a certain demographic, or as a certain marginalized group, and it almost feels like a caricature of that group.}'' \looseness=-1

    Taking this argument one step further, E22 argued that this misrepresentation of minority groups stems from generated data often defaulting to normative assumptions embedded within the auxiliary model. This raises a broader concern that the use of synthetic data reinforces cultural hegemony---that is, the dominance of one cultural framework (often Western) over others---by embedding dominant values in AI systems, perpetuating a worldview that marginalizes alternative cultural perspectives. Addressing such concerns would demand a fundamental reevaluation of the social and cultural assumptions underlying synthetic data generation.
    
\subsubsection{Organizational Pressure to Scale.}
While practitioners acknowledged that synthetic data has its limitations, many expressed a sense of resignation toward addressing them. Practitioners described how they were navigating a fast-paced and competitive AI landscape where organizational priorities often emphasized scaling data generation to meet the demands of model development and evaluation. While practitioners recognized that synthetic data could introduce biases or fail to capture the complexity of real-world data, they frequently proceeded despite these risks, with the intention of addressing potential issues later. P18 recognized this recurring tension:
    ``\textit{Frankly, this is a very busy space. I would love to take things slower, but it's just-- I don't know. We've had multiple failed attempts to do better documentation of our synthetic data, have better practices, which did not go very well with how fast we worked honestly.}''  
    
RAI experts shared concerns about a growing complacency toward addressing these limitations. E21 expressed a worry that the ``\textit{ease of use of synthetic data will so far outweigh, according to the calculus of certain groups, the harms that it does and they're just like `Oh, we'll just sort of know this is a limitation.'}'' Most participants noticed a tendency to rely on partial measures and surface-level fixes (e.g., prompt variation) rather than developing substantive longer-term solutions or foregoing the use of synthetic data. 
\looseness=-1

\section{Discussion}
\label{sec:discussion}
The ubiquitous use of auxiliary models reflects practitioners' optimistic perceptions that synthetic data can be a solution for challenges like data scarcity, privacy concerns, resource constraints, and high data collection costs. However, the anticipated benefits of synthetic data were seldom fully realized in practice.  The troubling prevalence of logics of scale and automation \cite{hanna2020against} signal an industry-wide prioritization of efficiency that positions synthetic data as the most practical---even if imperfect---option.  Indeed, many practitioners acknowledged unresolved challenges with their use of synthetic data and warned against the danger of normalizing these practices without critically assessing their implications for specific use cases and domains.  Practitioners described how the pressure to meet immediate demands often outweighed the desire for rigorous validation.  The scale and opacity of synthetic data created through auxiliary models further compound these problems as data issues become more challenging to trace, deferring accountability \cite{widder2023dislocated} and leaving systems more prone to perpetuating harm. Thus, the scale of synthetic data emerges as both a promise and a peril.  \looseness=-1

While our findings highlight significant concerns about the role of auxiliary models in development practices, we resist framing synthetic data as inherently problematic. An AI development pipeline that excludes synthetic data may involve other challenges, for instance, in meeting the ever-increasing demands of evaluation for robust model development across a range of harm areas and specialized domains, as required by policymakers or industry standards. We contend that the impacts of synthetic data are multifaceted and plural, with both potential benefits and risks, and invite critical, ongoing investigation by the FAccT community. \looseness=-1 

\subsection{Interrogating Impacts to the AI Supply Chain}

Our findings capture the perspectives of practitioners embedded within systems of institutional power. This `studying up' approach \cite{nader1972up, barabas2020studying} may fall short in surfacing how synthetic data use reconfigures other aspects of the AI supply chain \cite{widder2023thinking}. Below, we outline opportunities to investigate impacts on model deployers, data workers, and model subjects. \looseness=-1 

Building AI from scratch requires vast amounts of high-quality data that raises the barrier for entry for new companies seeking to develop applications \cite{widder2023open}. One might argue that synthetic data could contribute to the democratization of model development since smaller companies no longer need to have access to large-scale, costly, hard-to-obtain datasets to train their models \cite{lee2024synthetic}. In theory, synthetic data could counteract the concentration of power within the hands of a few incumbents by allowing startups or other entities to compete against large tech companies. 

Nevertheless, there is a paradox in this democratization narrative. The auxiliary models typically used to generate synthetic data are almost exclusively developed by the same tech giants that already dominate the AI industry \cite{widder2023open}. Smaller companies or researchers who wish to train a model themselves are likely to rely on these auxiliary models offered as-a-service to produce synthetic data \cite{luitse2021great}. This results in a consolidation of power as big tech companies monopolize the tools and frameworks \cite{burkhardt2024foundation} that smaller companies must rely on for generating synthetic data \cite{azoulay2024old}. As synthetic data pipelines become entangled with proprietary model APIs, usage restrictions, and tiered access to compute, the ability to experiment or contest how synthetic data is generated is increasingly constrained.

We must also consider how these emerging practices impact the data supply chain. In recent years, data work has shifted from independent platform workers to private annotation firms specializing in labeling services \cite{wang2022whose}. These firms often operate within opaque supply chains that hide the outsourcing of work to the Global South \cite{miceli2020between}. Participants noted that synthetic data was often used to avoid the need to engage with data workers, but training and testing infrastructures that use auxiliary models still require human input. The effects of the shift on labor infrastructures remain underexplored. How do these emerging practices impact job opportunities in data work? How does the increasing reliance on synthetic data potentially reshape data workers' roles from creators to `custodians of data quality'? How might synthetic data practices provide opportunities for data workers' upward mobility, and under what conditions would such shifts be equitable? These questions require further investigation.   \looseness=-1

On the other hand, auxiliary model use offers potential opportunities to mitigate some of the harms associated with traditional data annotation, particularly in cases where workers are exposed to toxic or harmful content. Participants mentioned how automating the evaluation of certain types of harms (\textbf{e.g.,} those involving violent, graphic content) and relying on annotators for subtle or complex forms of harm, could reduce the need for human annotators to engage directly with harmful material that has been linked to psychological distress and trauma \cite{steiger2021psychological}.  \looseness=-1

Finally, the use of synthetic data for training and evaluating models has significant implications for model subjects. Studies have shown that distribution shifts caused by synthetic data during training can lead to reductions in model performance and fairness, and to increased representational harms for minoritized groups \cite{wyllie2024fairness}. If auxiliary models are \textit{also} used in the evaluation, they might fail to detect critical issues, particularly those that disproportionately affect underrepresented groups or rare phenomena. This risk of degraded model performance during training, which is likely unnoticed in evaluation, also highlights the urgency of advancing research in this area. What mechanisms could then be developed to scrutinize and mitigate the ways in which general-purpose models propagate biases into synthetic datasets?  \looseness=-1

\subsection{Toward Considerations for Responsible Use of Synthetic Data}

Our interviews reflect a particular moment in time when practices and norms around the use of synthetic data in both industry and research contexts are rapidly evolving. The viewpoints expressed by different participants were often in tension with each other or with the growing synthetic data literature \cite{qian2024evolution}. Some participants argued that synthetic data is well-suited for use in training due to its tolerance for noise but is too risky to use in evaluation, while others expressed the opposite viewpoint. Some praised the use of synthetic data because it gave them ``\emph{full control},'' while others recounted difficulties controlling data generation in practice. Some proposed that auxiliary models can be used to simulate data from members of underrepresented groups, a viewpoint some literature supports~\cite{grossmann2023ai}, while others argued that data generated in this way would be merely a ``\emph{caricature}'' of the group, in line with other scholarship~\cite{Wang2024llms}. These disagreements reflect the contested nature of our collective understanding of synthetic data's benefits, limitations, and appropriate use. Because of this, rather than offering prescriptive recommendations, we focus on foregrounding key considerations that researchers and practitioners must navigate as they engage with synthetic data.

\textbf{Practitioners should carefully consider what constitutes an appropriate application of synthetic data.} Deciding whether to use synthetic data, and in what capacity, involves assessing the intended purpose or use case and the domain of application, and balancing trade-offs between resource availability and the `right' methodology for data generation and validation. For example, in a `high-risk' domain (like training a model for a medical application), it is arguably more crucial to clearly operationalize the desiderata (diversity, for instance) and to systematically validate that the data meets those characteristics. In `low-risk' domains, there is perhaps more tolerance for noise or fuzzy data. Future research could investigate which dimensions of a use case or domain are relevant in determining the appropriate scope of use for synthetic data.  Of course, what counts as high-risk or low-risk is itself a value-laden assumption that must be interrogated.

\textbf{Practitioners should carefully consider auxiliary model capabilities and limitations when selecting which model to use. }Synthetic data practices are heavily entangled with the selection of auxiliary models. The harms of synthetic data are a result of the assumptions and design decisions made during the creation of synthetic data \textit{and} the limitations of the auxiliary models. Each model has a different risk profile \cite{uuk2024effective}, which makes auxiliary model selection a consequential decision. Despite this, participants in our study often defaulted to selecting what they perceived as the state-of-the-art model.  This tendency reflects a broader pattern in AI practices, where newer models are assumed to outperform older ones \cite{newerornot}, even when their suitability for particular contexts remains untested.  Practitioners could benefit from a more deliberate approach to model selection, where they experiment with whether a particular model aligns with their specific needs. \looseness=-1

\textbf{In addition to careful model selection, there is a need to make synthetic data validation more systematic.} Participants in this study relied on `eye-balling' or `spot-checking' as their primary validation method. While manual review offers a direct way to identify errors, participants frequently reported it as inconsistent, insufficient, or deprioritized over the demands of data generation. First, practitioners need to invest more time and resources in synthetic data validation. Second, practitioners should develop a methodology that ideally combines multiple forms of data validation (e.g., manual verification \textit{and} calibration experiments). There would be several considerations to keep in mind. For example, instead of randomly sampling data, one could prioritize defining (including with the absence of real-world data) and creating underrepresented or edge-case examples. Similarly, to guide validation efforts, practitioners could develop a rubric that articulates the criteria against which they compare the synthetic data. Yet, how can such qualities be effectively measured, and how might reliance on proxy metrics introduce additional risks? We call for research and experimentation to develop effective validation practices for synthetic data. \looseness=-1 

\textbf{Researchers should also support the development of artifacts and processes for synthetic data documentation.} Data practices among our participants were highly iterative and typically involved multiple rounds of generation and validation, interleaved together, making synthetic datasets fluid and iterative objects that change constantly. However, rarely did participants prioritize documentation of data or process, either due to organizational priorities or the perception that code or prompts served as sufficient documentation for synthetic data. While there are plenty of resources for data and model documentation~\cite{mitchell2019model,gebru2018datasheets,bender2018data,cleardoc}, there is almost no existing guidance on what to document for synthetic data that is created and used in such myriad ways. The EU's General Purpose AI Code of Practice \cite{euaicodeofpractice} recommends documenting the methods used to generate synthetic training data, but does not yet get into specifics or touch on use cases beyond training. What aspects of the synthetic generation process should be recorded? For example, how should revisions to prompts, examples of rejected data, or the reasoning behind design decisions be captured? Practitioners should also consider documenting their choice of the auxiliary model as well as a justification. Beyond recording data provenance, there are broader questions about synthetic data documentation that warrant attention: Who are the intended audiences, and what purposes should synthetic data documentation serve? How should documentation practices be adapted to account for the experimental and iterative process that practitioners described?\looseness=-1

\textbf{Lastly, AI workflows that incorporate auxiliary models must prioritize meaningful community engagement.} Synthetic data inherently reduces opportunities for direct participation, as it is neither created nor annotated by individuals or communities. Often, an underlying motivation for using synthetic data is to ``take the people out'' \cite{susser2024synthetic}, whether to address privacy concerns, meet compliance requirements, or minimize exposure to harmful content. While these intentions may serve some goals, they also exclude data subjects and data workers from the pipeline and restrict opportunities to question or shape datasets. To increase participation, participants in our study proposed a consultation model where experts or community members provide input on use cases, evaluation outcomes, or critical decisions \cite{delgado2023participatory}. However, when is it appropriate to consult only experts, and when should affected communities play a central role? The answer likely depends on the specific use case and the degree of transformation involved. For example, small or targeted modifications to existing datasets might raise different participatory needs compared to entirely synthetic datasets. The timing of participation is equally significant. Should stakeholders contribute during the generation process, shaping decisions about what data is created and how? Or is it more effective to involve them during validation? These decisions shape the accountability of synthetic data practices to the communities they may impact.  \looseness=-1

In highlighting these considerations, we recognize how meaningful change requires rethinking incentives for different stakeholders across the AI supply chain. First, regulation could establish reporting requirements for both auxiliary model producers and downstream developers. For model producers, this could involve disclosing evaluations of their models' capacity to generate synthetic data in specific domains, along with guidance on the intended scope of use. Developers using synthetic data could be expected to document prompts, generation parameters, and any modifications made to the outputs. Furthermore, demonstrating that this kind of reporting leads to better outcomes (e.g., product usage or increased user trust) would establish clear incentives for these evaluations. 

\section{Conclusion}

In this paper, we illustrate the expanding role of synthetic data in AI development.  Through interviews with AI practitioners and RAI experts, we highlight the motivations and promises of synthetic data, its various use cases across the AI development pipeline, current practices and challenges around its generation and validation, its limitations, and ethical considerations for its use. We discuss the implications of our findings for the AI supply chain. Finally, we propose several considerations for improving synthetic data practices. We note that while synthetic data is valued precisely for its efficiency, practices like rigorous validation, documentation, or participatory methods all require human intervention, increasing costs and extending timelines. To arrive at adoptable practices, we must reconcile competing demands between the need for rigor and the desire to maintain efficiency in line with organizational priorities and constraints.\looseness=-1


\section{End Matter}
\label{sec:ethics}

\textbf{Ethical considerations.} We took several measures to protect our research participants. Our study design was reviewed and approved by our organization's Institutional Review Board. Participation was voluntary. Informed consent was obtained electronically and we compensated participants with a gift card of \$75 USD.  Participants were informed of the purpose of the study and were told that they could refuse to answer any questions and ask for the recording to be paused at any time.  All participant data, including interview transcripts and recordings, was stored in a secure location and was not accessible by anyone outside the research team.  The research team checked all included quotes to ensure that no identifying information about participants or their organizations was revealed. \looseness=-1

\textbf{Positionality.} All authors are, or were at the time of conducting the research,
employed at a large technology company. This positioning afforded us access to ongoing conversations within the industry around synthetic data practices, including emerging use cases. At the same time, we recognize that our proximity to industry priorities may shape how we frame risks and recommendations.  
Throughout this work, we have aimed to remain accountable to perspectives that may not align with institutional interests by foregrounding scholarship from critical data studies and the political economy of the AI supply chain. We offer this reflection to acknowledge how our embeddedness within industry shapes the kinds of questions we ask and the interpretations we bring to our analysis, and to invite ongoing dialogue with communities who approach this topic from different positions or commitments.


\textbf{Potential adverse impacts.} First, our findings focus exclusively on practitioners' and RAI experts' perspectives. Our findings may inadvertently privilege their viewpoints and concerns over
those of other crucial stakeholders. This could lead to recommendations that undervalue the needs of affected communities, policymakers, and civil society organizations. 
Second, our research sheds light on motivations, practices, and challenges for the use of synthetic data in the AI development pipeline.  We also expose limitations and ethical considerations.  Emphasizing these drawbacks may be interpreted as evidence that the use of synthetic data should be more restricted.  However, we caution against this interpretation of our results.  In setting policy guidance for synthetic data, policymakers should weigh the benefits as well. Perhaps most notably, synthetic data enables the evaluation of models and systems at a scale that would not be possible through other means; without it, many evaluations simply may not happen. To address this concern, we have aimed throughout this paper to provide a balanced perspective, highlighting both the benefits and drawbacks of synthetic data.  We additionally emphasize that norms and practices around the use of synthetic data are rapidly evolving.  While we hope that our research inspires future work developing best practices for the responsible use of synthetic data, these practices will need to be constantly evaluated and updated to ensure that they remain practical and relevant. 
\begin{acks}
First and foremost, we are grateful to our study participants, without whom this research would not be possible.  This work was enriched by many useful conversations with Anubhav Jangra, Nari Johnson, Daniela Massiceti, Cecily Morrison, Sachita Nishal, and the members of the Microsoft Research FATE group, Microsoft's Sociotechnical Alignment Center, and the Tech Solidarity Lab at CMU.  Our sincere thanks to Hanna Wallach and Ece Kamar for making connections to the AI practitioner community and helping with recruitment. We are also grateful to Agathe Balayn, Inha Cha, Sarah Fox, and Jordan Taylor for providing feedback on early drafts of this paper. 
\end{acks}

\bibliographystyle{ACM-Reference-Format}
\bibliography{refs}

\appendix
\section{Additional Study Details}
\label{sec:appendix}
In this appendix, we provide the interview protocols used in each phase of our study.

\subsection{Interview Protocol for Phase 1}
\textit{Note:} This interview protocol contains questions that are not relevant to some participants. The moderator decided which question to follow up with based on the participant's AI pipeline and project description. For example, participants whose focus was on evaluation were not asked questions related to training, and vice versa.

\subsubsection{Introduction}
\begin{itemize}
    \item Introduction (role, background, years of experience)
    \item How large is your team? 
    \item What kinds of projects do you focus on? 
    \item What types of data do you deal with? 
\end{itemize}

\subsubsection{ML pipeline}
Now would be a good time to shift gears and learn a little more about a specific project where you were involved in data generation, curation, or model evaluation.
\begin{itemize}
    \item Could you give me a brief overview of your ML pipeline step-by-step, from data development to model training, evaluation, and deployment? It would be helpful to get a full picture of the various stages where you deal with different kinds of data.
    \item Follow-up clarification questions:
    \begin{itemize}
        \item Are there any labels involved in this data? 
        \item What are your data sources? Where does this data come from?
        \item Do you augment the dataset in any way or make perturbations?
        \item Would you say [data they described] is synthetic data? Why or why not?
    \end{itemize}

\end{itemize}

[If they do not already talk about model-generated or simulated data as part of the earlier question]
Have you ever or do you currently use synthetic, simulated, model-generated, or augmented data in your work? For example, data generated by LLMs that you use to train or evaluate other LLMs, or simulating users with LLMs by automatically generating prompts.
\begin{itemize}
    \item What do you usually call this kind of data?
    \item Would you consider this synthetic data?
    \item Are there other ways you have worked with synthetic, simulated, model-generated, or augmented data?
    \item What counts as synthetic data in your field? What terms do you use to describe this kind of data?
    \item (If they say no to all forms of synthetic data) Did you ever consider using these kinds of data but then decide against it? If so, why?
    \item Do you ever use models to evaluate other models or systems? For example by simulating users or automatically generating prompts? What made you try this approach?  
    \item Do you also incorporate human feedback into their evaluations?  
    \item How do you decide when you need humans in the loop and when you don't?
    \item Can you walk me through your process of generating synthetic data in some detail?
\end{itemize}
               
\subsubsection{Model training}

\begin{itemize}
    \item Do you use synthetic data exclusively for training, or is it combined with real-world data?
    \item Are there specific things you need to do or look out for before you combine synthetic data with other data sources, such as real world data? 
    \item Have you encountered any performance bottlenecks or computational challenges when working with large-scale synthetic datasets? 
    \item Do you have concerns relating to distribution shifts or `model collapse'?
\end{itemize}

\subsubsection{Model evaluation}

\begin{itemize}
    \item What kinds of evaluation did you run for this project?
    \item What is the goal of the evaluation? Do you have specific objectives for the evaluation?
    \item Do you incorporate feedback from evaluations with synthetic data into your model development process? If so, how? 
\end{itemize}

\subsubsection{Prompt-generated data} 

\begin{itemize}
    \item How did you choose the pre-trained LLM? What were the factors that went into this decision?
    \item How do you come up with the prompt templates that are used to generate data with the LLM? (Is this a manual or automated process?) What kinds of expertise is needed to select the prompt templates?
    \item Could you share an example of a prompt that you used? Did you try different variations of prompts before deciding on this approach? When evaluating data for each iteration of the prompt, what were you assessing in the data? Why wasn’t it meeting your needs or what led you to changing the prompt?
    \item What makes this technique effective for your use case?
\end{itemize}

\subsubsection{Understanding synthetic data practices}
\begin{itemize}
    \item Did you always use synthetic data in your data pipeline, or did you make a shift? If so, why?  
\item How did you decide to use synthetic data?
\item What makes synthetic data appealing for the work you do? [to get at all the benefits]
\item What objectives does it serve for you? What goals do you have?
\item Did you face technical or organizational challenges when shifting to synthetic data?
\item What does good synthetic data look like? 
\item How do you define success for synthetic data?
\item What counts as high-quality synthetic data for your work?
\item Do you ever think about the ‘realistic-ness’ of synthetic data? How ‘realistic’ does synthetic data need to be for your purposes?
\item How do you think about glitches, artifacts and imperfections of real-world data?  
\item How do you evaluate or validate synthetic data? Fidelity, reliability etc.
\item For evaluating the generated data: do you qualitatively look through the samples or use any metrics?
\item Where do you face the biggest challenge with data?
\item Have you deployed a system that uses synthetic data? 
\item Have you ever noticed any downstream issues that you attribute to synthetic data?
\end{itemize}

\subsubsection{Responsible AI considerations}
\begin{itemize}
    \item Are you aware of any responsible AI considerations with the use of synthetic data? 
    \item How do you navigate these concerns? 
    \item Have you come across any guidance or best practices for using synthetic data?
    \item Are there any resources that you found helpful?
    \item Do you think synthetic data helps you mitigate responsible AI issues (such as privacy, fairness-related harms, or consent)?
    \item On the other hand, does it also exacerbate any responsible AI issues?
    \item Transparency:
    \begin{itemize}
        \item Do you record any information or create documentation for synthetic data?
        \item What kinds of information do you record? 
\item Would you say that it is similar to ‘real-world’ data or are there fields unique to synthetic data that you need to include?
\item Is this data shared publicly or used by other teams? Do you have to process the data to make it usable for other teams?
\item 
  When you are using model-generated data, does the data underlying the model inform your approach? Do you have a good sense of the data that is used to train the model you are using? 
\item Does the provenance of synthetic data matter for the work you do?
    \end{itemize}
    \item Fairness:
    \begin{itemize}
        \item How do you think about fairness in the context of synthetic data?
        \item Is there any form of ‘debiasing’ that needs to happen with synthetic data? 
        \item Are there considerations around diversity and representation that are relevant to synthetic data? 
    \end{itemize}
\end{itemize}

Are there things you would like to share that we did not get to discuss today? 

Thanks for your time!

\subsection{Interview Protocol and Vignettes for Phase 2}

\begin{table*}[htbp]
\centering
\small
\renewcommand{\arraystretch}{1.5} 
\begin{tabular}{>{\raggedright\arraybackslash}p{4.5cm} | >{\raggedright\arraybackslash}p{4.8cm} | >{\raggedright\arraybackslash}p{4.5cm}}
\textbf{Scenario 1:} & \textbf{Scenario 2:} & \textbf{Scenario 3:} \\
\textbf{Fine-tuning data} & \textbf{Evaluation} & \textbf{Simulated data} \\
\hline
The SmartTeach team is developing question-answering (QA) systems for tutoring. They would like to offer hints within their system to scaffold learning. 
Due to limited task-specific data, the team is leveraging a pre-trained Language Model (LLM) to generate hints for an existing QA dataset. The hints dataset will be used for fine-tuning a smaller generative model. &
The Responsible AI team works on identifying harms within LLM conversations. People need to interact with an LLM to understand failure modes, but this is often expensive to scale and is also time-consuming. The Responsible AI team is exploring a new technique: they are prompting an LLM with a topic \& user identity characteristics. This LLM will have multi-turn conversations with the system-under-test. Following these interactions, another LLM ‘scores’ the conversation to detect potential harm. &
The PrivacyOps team supports internal teams to better understand user feedback while safeguarding data privacy. They are using an LLM to simulate employee interactions within a hypothetical organization. This involves creating personas and generating synthetic documents and emails that resemble real-world situations. This simulated data will be used to test models and train downstream classifiers. \\
\bottomrule
\end{tabular}
\caption{Vignettes of Synthetic Data Use \label{tab:vignettes}}
\end{table*}

\subsubsection{Introduction}
\begin{itemize}
    \item Introduction (role, background, years of experience)
    \item What does responsible AI mean for the work you do? Are there any specific areas of RAI that you focus on?
    \item What kinds of projects do you work on?
\end{itemize}

This is a good time to shift gears and discuss the recent shifts towards using synthetic data or model-generated data or simulated data or augmented data in the ML pipeline (e.g., for data creation or model evaluation), and how that might impact responsible AI considerations. 

Before we go into specific scenarios of how this data is being used today, it would be nice to start off hearing your broad perspective on these emerging practices.  
\begin{itemize}
    \item Has your work touched on the use of synthetic, simulated, model-generated, or augmented data across the ML pipeline or have you previously thought about the use of such data?
    \item Are you aware of any responsible AI considerations with the use of synthetic data? 
    \item Do you think synthetic data helps you mitigate responsible AI issues (such as privacy, fairness-related harms, or consent)?
    \item On the other hand, does it also exacerbate any responsible AI issues?
\end{itemize}

Thanks, this is very helpful. Here are the three scenarios. [Show slide with the three scenarios presented in Table~\ref{tab:vignettes}.]

\subsubsection{Reflecting on vignettes}
For each vignette:
\begin{itemize}
    \item What are the major concerns of using models for this purpose?
    \item What kinds of trade-offs are involved in this scenario and are there good ways to make decisions around this?
    \item Do you think there is a need to validate the generated data? If yes, which properties might be important to validate?
    \item What are the implications of not doing this validation?
    \item What kinds of strategies might help us validate (evaluate) synthetic data or model-generated data?
    \item Do you anticipate any potential risks or downstream impacts with using model-generated data in this way?
\end{itemize}

Thanks for covering these scenarios with me.

\begin{itemize}
    \item What kinds of guidance or best practices should teams follow when creating/using synthetic data?
    \item Are there contexts where using model-generated data might be better than traditional methods? Why?
    \item We know of trade-offs around synthetic data (e.g., privacy and utility, realistic data \& copying training data). How should practitioners navigate these trade-offs and make decisions on when \& how to use synthetic data?
\end{itemize}

\subsubsection{Responsible AI considerations}

\begin{itemize}
    \item Does synthetic data mitigate some responsible AI issues? 
    \item Does synthetic data also exacerbate some responsible AI issues? 
    \item Does this use of synthetic data introduce new challenges or directions for responsible AI work?
    \item How can synthetic data help or hinder transparency in AI systems?
    \item How does the process of creating synthetic data impact ethical considerations?
    \item Do we need to think about fairness differently in the context of synthetic data?
    \item Are there considerations around diversity and representation that are relevant to synthetic data?   
    \item How should the provenance of synthetic data be documented to ensure accountability and trustworthiness?

\end{itemize}

What roles do stakeholders (such as model producers, researchers, regulators, and civil society) play in ensuring responsible use of synthetic data across the ML pipeline?
 
Are there things you would like to share that we did not get to discuss today? 

Thanks for your time!





\end{document}